# Cluster Ellipticals at High Redshift: The View from the Ground and with HST

Mark Dickinson

*STScI, 3700 San Martin Dr., Baltimore, MD 21218*

**Abstract.** New ground–based and *HST* observations of distant clusters make it possible to trace the history of E/S0 galaxies to lookback times $\sim 10 h_{50}^{-1}$ Gyr. The data strongly favor a scenario in which cluster ellipticals formed very early with a narrow spread in ages. By $z = 1.2$ there are changes in color and luminosity consistent with simple passive evolution, but even at that redshift the galaxies appear mature, with red colors suggesting ages of several Gyr. Apparently we are still falling well short of seeing the true epoch of formation.

## 1. Introduction

Much of the excitement over galaxy clusters at high redshift has focused on the apparently dramatic evolution of their galaxy content. Today, the cores of rich clusters are dominated by elliptical and S0 galaxies; the few spirals which are present are generally red, HI–deficient, and apparently quite inert. At higher redshifts, however, the situation seems substantially different, with a population of blue cluster members found even in the central regions of quite rich systems.

Here I would instead like to concentrate on the somewhat neglected, seemingly duller, but dominant cluster members: the early–type galaxies. The traditional view of giant elliptical (gE) galaxies has them forming at high redshift with a single, major episode of star formation, followed thereafter by a long period of simple, passive evolution. Ellipticals obey a tidy color–magnitude (c–m) sequence, with smaller, fainter galaxies exhibiting bluer colors. This is generally interpreted as a mass–metallicity relation. The small dispersion in gE colors around the mean c–m relation has been cited as evidence that the galaxies formed quickly and within a time interval that is short relative to their present ages. Bower *et al.* (1992) have compared the color–magnitude sequences of E/S0 galaxies in the Coma and Virgo clusters, concluding that the galaxies in those two clusters at least are uniformly old and coeval, with little residual star formation at later times.

This simple, traditional picture of ellipticals has been challenged. At this meeting, detailed spectral analyses of age and metallicity indicators were presented which suggest that ellipticals may have formed over a broad interval of cosmic time, with many galaxies displaying signs of substantial star formation as recently as a few Gyr ago (cf. Worthey *et al.* in this volume). In addition, careful scrutiny of ellipticals often reveals shells and other non–uniformities in their light profiles which probably result from the ingestion of other galaxies.

This has led to the extreme alternative viewpoint that many ellipticals formed at late times as the byproducts of major mergers between disk galaxies. Numerical simulations show that the products of such mergers resemble elliptical galaxies, with spheroidal stellar distributions and $r^{1/4}$–law light profiles.

Much of the evidence called to bear in this debate has been derived from observations of nearby galaxies. It is possible, however, to peer back in time, studying the properties of ellipticals at high redshift. Galaxy clusters serve as our laboratories, as they are elliptical–rich and identifiable out to large redshifts. A variety of new observational tools are being brought to bear on distant clusters, most notably deep images with high angular resolution provided by the *Hubble Space Telescope (HST)*. At $z = 0.5$, the $0''\!.1$ resolution of *HST* provides images equivalent to observing the Coma cluster in $1''$ seeing! We need no longer infer what sort of galaxies we are observing from colors or spectra – we can simply choose E/S0 galaxies morphologically at whichever redshifts we can find them.

In order to disentangle the evolution of the dominant stellar population from small contributions of later star formation, it helps to measure galaxy colors in the near–infrared. The long–wavelength spectra ($\lambda \gtrsim 7000$Å) of mature galaxies are dominated by light from old, evolved stars, primarily red giants, and are minimally subject to perturbation by recent star formation. Bluer colors are sensitive to hotter stars, reflecting the main sequence turn–off for the dominant population plus any contribution from later star formation. For a passively evolving, single–burst model galaxy, a rest–frame color like $(V - H)_0$ reddens rapidly during the first few Gyr while hot stars provide a substantial fraction of the total luminosity, and then levels off, evolving very slowly thereafter. Conversely, looking backward in time toward higher redshifts, we may expect optical–IR or IR–IR colors of our cartoon passively–evolving elliptical to remain red until the galaxy age drops below ∼3–4 Gyr, at which point they should plummet blueward. At high redshift, photometry at rest–frame wavelengths $\lambda_0 \gtrsim 7000$Å requires infrared measurements.

## 2. Cluster ellipticals at z < 1

The first studies of distant clusters using IR arrays were presented by Aragón–Salamanca *et al.* (1991,1993), who observed a sample of distant, optically–selected clusters. They found that the $V - K$ and $I - K$ colors of cluster galaxies exhibit a prominent red sequence (presumably the E/S0s), and that at high redshifts this sequence shifts toward bluer colors. At $z \approx 0.9$ they measured few, if any, galaxies as red as present–day ellipticals. New large–format array cameras make it possible to go deeper faster, and over a much wider area. Adam Stanford, Peter Eisenhardt and I have been carrying out such a survey, obtaining deep, wide–field images of a large number of clusters spanning as wide a redshift range as possible. We obtain data at $J$, $H$ and $K$, as well as in two optical bands chosen to bracket $\lambda_0 4000$Å at the cluster redshift. The images typically cover a field of view ∼2 Mpc on a side at the cluster redshift, and reach 2 mag below present–day $L^*$ with $S/N > 5$. To date, we have observed more than 40 clusters, ranging from nearby Coma out to $z = 0.9$, the redshift limit of existing cluster catalogs.

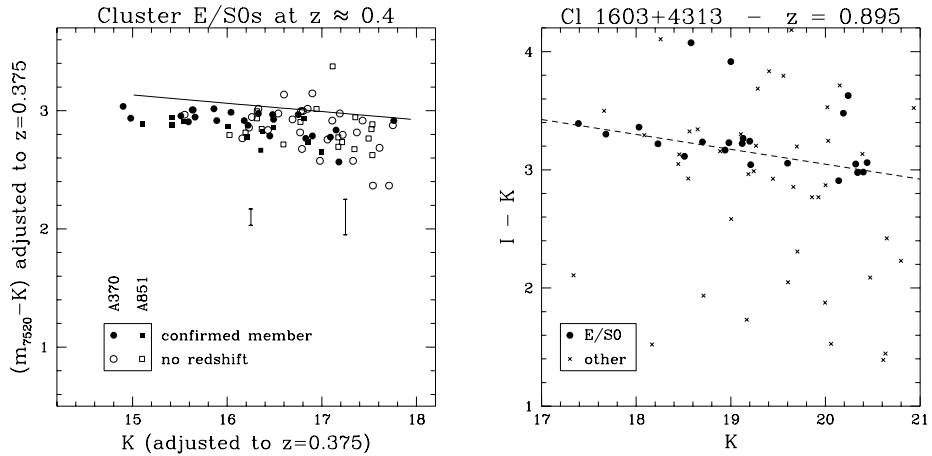

Figure 1. *Left:* Optical–IR color–magnitude diagram for E/S0 galaxies in Abell 370 ($z = 0.375$) and Abell 851 ($z = 0.407$), with *HST* morphological classifications provided by A. Oemler. The A851 photometry has been adjusted to match the redshift of A370. Representative $\pm 1\sigma$ error bars are shown. The solid line traces the mean $(V - H)$ vs. $H$ c–m relation for E/S0s in Coma as it would appear if redshifted to $z = 0.375$.

Figure 2. *Right:* Color–magnitude diagram for Cl 1603+4313, a cluster at $z = 0.895$ observed with *HST* by Westphal. The dashed line is a fit to the E/S0 c–m relation.

Figure 1 is adapted from Stanford, Eisenhardt and Dickinson 1995, and shows an optical–IR color–magnitude diagram for galaxies in Abell 370 and Abell 851 (a.k.a. Cl 0939+47), two clusters at $z \approx 0.4$. Both clusters have been observed by *HST* (Couch *et al.* 1994, Dressler *et al.* 1994), and figure 1 plots only those galaxies classified as E/S0. At this redshift, our observed bands translate neatly to rest frame $V$ and $H$, requiring only small differential k–corrections to make the match exact. The solid line shows a least–squares fit to the $(V-H)_0$ vs. $H$ color–magnitude relation for Coma cluster ellipticals from Bower *et al.* (1992 and priv. comm.). Here no reference to models is needed: we simply choose E/S0s in both Coma and A370 from their morphologies, measure colors in the same rest–frame bands, and compare.

The mean Coma c–m relation forms a red boundary to the distribution of the high redshift E/S0s. There appears to be a slight shift ($\Delta(V - H)_0 = -0.13$ and $-0.18$ mag for A370 and A851, respectively) toward bluer colors compared to Coma. However this shift is only marginally significant when possible sources of systematic error are taken into account. These include seeing effects, aperture and color gradient corrections, the (small) k–corrections, and uncertainties in the photometric calibrations. Together, these result in $\sim 0.06$ mag uncertainty in the color comparison, making the apparent bluing for A370 and A851 $2\sigma$ and $3\sigma$ effects, respectively. The point is that color evolution, if present at all, is very small – 0.1 to 0.2 mag over a baseline spanning a factor of 3 in wavelength. This is quite consistent with passive evolution models which match today's gE's with ∼13 Gyr–old stellar populations. The slope and scatter of the c–m relation also match those found today. The slope is $d(V - H)_0/dH \approx 0.06 \pm 0.02$ mag per mag, compared to $0.07 \pm 0.01$ found by Bower *et al.* for Coma. For galaxies brighter than $L^*$, where photometric errors are small, the intrinsic scatter in the $\Delta(V - H)_0$ color (i.e. the color difference after removing the slope of the c–m relation) is approximately 0.06 mag. Again, this is very similar to the value of 0.05 mag measured for Coma by Bower *et al.* Considering that $\sim 5h_{50}^{-1}$ Gyr have elapsed since $z \approx 0.4$ (where $h_{50} = H_0/50 \,\mathrm{km\,s^{-1}\,Mpc^{-1}}$), the formation epoch of cluster ellipticals must lie substantially prior to that redshift.

Figure 2 shows a similar c–m diagram for Cl 1603+43 at $z = 0.895$, also observed by *HST*. E/S0s, which I have classified from the archival WFPC2 images, are marked as circles, while all other galaxies are shown as crosses. No redshifts have been published for individual galaxies, so the plot presumably includes both field galaxies and cluster members. However the c–m sequence is easily visible as a narrow swath of E/S0s. Discarding some faint, very red outliers which could well be background galaxies, the c–m slope is 0.12 mag per mag, close to the $z$=0 value for rest–frame $(B - J)$ colors. The scatter is again small, < 0.07 mag rms. It appears that even at a lookback time of 8 to $9h_{50}^{-1}$ Gyr, ellipticals were highly coeval within this cluster. The fact that the color scatter is not measurably increasing indicates that the fractional dispersion in galaxy ages was small even then, while the stability of the c–m slope implies that cluster ellipticals have experienced little differential evolution (i.e. as a function of luminosity) in either age or metallicity since $z \approx 0.9$.

## 3. Cluster ellipticals at z > 1

Optical and x–ray samples of galaxy clusters peter out by $z \approx 0.9$. This may simply reflect the tremendous difficulty of finding a cluster and confirming its reality at higher redshifts, where galaxies are extremely faint and cluster contrast against the foreground/background of faint field galaxies is poor. At $z > 1$, the 4000Å break, a prominent spectral discontinuity in early–type galaxies, shifts through the $I$–band, and clusters beyond this limit might simply "disappear" optically. Hunting for clusters in the near–IR would solve this latter problem, but even today's arrays are not large enough to permit a blind, wide–field survey to the magnitude limits required.

In order to improve the odds, Peter Eisenhardt and I are conducting a directed survey, using radio galaxies as signposts to guide our telescopes. It has been known for some time that powerful radio sources (radio galaxies and radio–loud quasars) at $z \sim 0.5$ often inhabit clusters (cf. Yee and Green 1984, Ellingson *et al.* 1991, Yates *et al.* 1989, Hill and Lilly 1991, Allington–Smith *et al.* 1993, Dickinson 1994). We can hope that this holds at higher redshifts as well, where no clusters are currently known but samples of radio galaxies and quasars abound. The AGN serves as a profitable place to look, and also provides a likely redshift *a priori* for any cluster we may find. We have surveyed fields around $\sim 25$ powerful radio galaxies in the redshift range $0.8 < z < 1.4$, imaging deeply at $R$, $J$ and $K$. We compare galaxy counts and color distributions in each field to data from two large blank–field surveys which we have carried out with other collaborators. Cluster candidates are identified using two criteria: enhanced galaxy density on the sky at $K$, and/or unusual color distributions. If elliptical galaxies remain intrinsically red out to $z \approx 1$, their *apparent* $R - K$ and $J - K$ colors should k–correct to extremely red values (e.g. $R - K \gtrsim 6$). Any excess of such objects around a radio galaxy provides a strong hint of the presence of a cluster.

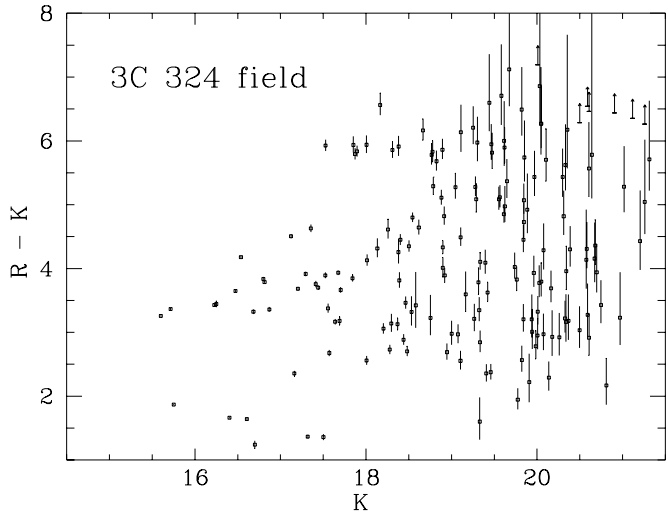

Figure 3. C–m diagram for the 3C 324 field, showing the narrow red sequence at $R - K \approx 5.9$.

The richest cluster candidate we have found to date surrounds the $z = 1.206$ radio galaxy 3C 324. A large excess of very red galaxies are concentrated around the radio source. Figure 3 presents a color–magnitude diagram of this field, showing the narrow red "finger" of galaxies with $R - K = 5.9$ extending faintward of $K = 17.5$. In Cycle 4 we used *HST* to image the 3C 324 cluster in order to study the morphologies of these galaxies. An 18 hour exposure with WFPC2 through the F702W filter confirms that they are, with few exceptions, the early–type galaxies we expected. Evidently, mature red ellipticals were already in place in rich clusters as long ago as $z = 1.2$. Figure 4 shows three examples, together with surface photometry demonstrating their $r^{1/4}$–law brightness profiles. I return to the structural properties derivable from the surface photometry in §4 below.

At $z = 1.2$ the observed $R$ and $K$ bands do not translate to well–studied rest–frame equivalents, so we cannot make the sort of direct comparison to Coma photometry that we used at $z = 0.4$. Relying on models of $z = 0$ ellipticals, and on the few empirical spectral templates available covering this wavelength range (e.g. Kinney *et al.* 1995), we find that the observed color is ∼0.6 mag bluer than the expected "no–evolution" value. This agrees qualitatively with the previous measurements by Aragón *et al.* at $z \approx 0.9$. Even at $z = 1.2$, however, the scatter in colors is remarkably tight. Restricting our attention to galaxies $K < 19$ (or $L_K \gtrsim L^*$ for ellipticals at this redshift) where photometric uncertainties are small, there are 13 galaxies in the red "finger" with an intrinsic rms color scatter of 0.20 mag. One galaxy contributes most of the variance, however. Morphologically an elliptical but 0.6 mag redder than the others, it lies ~1 Mpc from the cluster center in projection. If we allow ourselves the luxury of discarding it under the assumption that it may be in the background, the intrinsic scatter of the remaining 12 is only 0.07 mag rms – again very similar to the values found for clusters at lower redshifts.

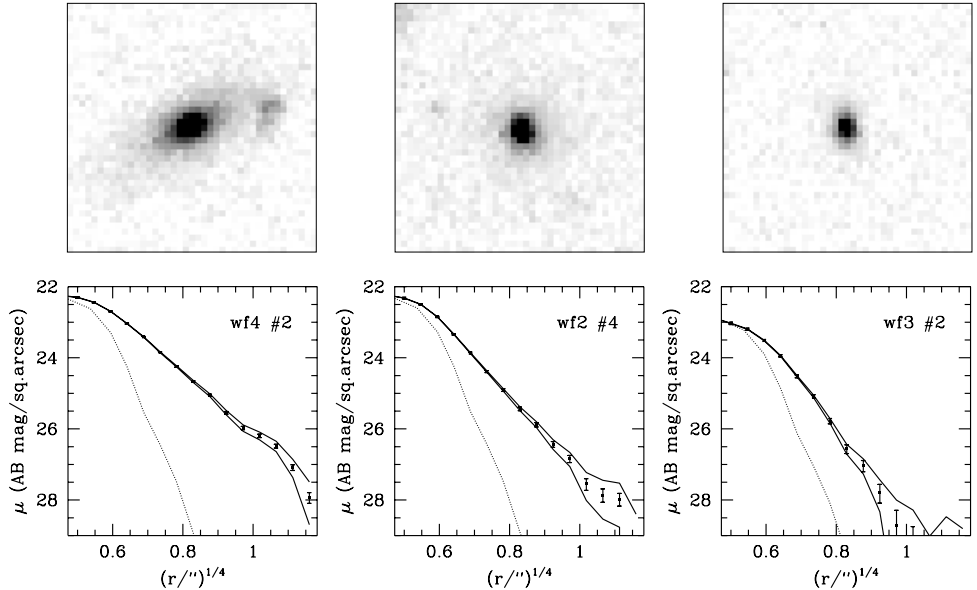

Figure 4. WFPC2 images and surface brightness profiles of three ellipticals around 3C 324. The images are 4″ on a side. The solid lines in the profile plots trace an envelope accounting for possible uncertainty in sky subtraction, while the dotted lines show the PSF. Note the companion or arm on WF4 #2, which is the 2nd–ranked cluster galaxy.

Age–dating a galaxy using broad–band colors is, at best, naive. Galaxies are composite stellar systems whose color is determined by age, metallicity, and the detailed stellar mix governed by their IMF and star formation histories. There also remain substantial uncertainties in many aspects of existing population synthesis codes (cf. Charlot, Worthey and Bressan 1995). Throwing caution to the wind, however, I cannot resist noting that a single–burst, solar metallicity Bruzual & Charlot model matches the $R-K$ color of the 3C 324 galaxies ~3 Gyr after the end of the burst. Given a lookback time of $\sim 10 h_{50}^{-1}$ Gyr to $z = 1.2$, this is nicely consistent with galaxy (and globular cluster) ages of ~13 Gyr today, provided of course that the Universe is old enough to accommodate such objects. Higher metallicities would permit younger ages. Adopting the method of Bower *et al.* (1992) to analyze the observed *scatter* in colors, I find that elliptical formation in the 3C 324 cluster must have been synchronized to within ~10% of the cluster age at $z = 1.2$, i.e. with $\Delta t \lesssim 300$ Myr assuming the age derived from the models.

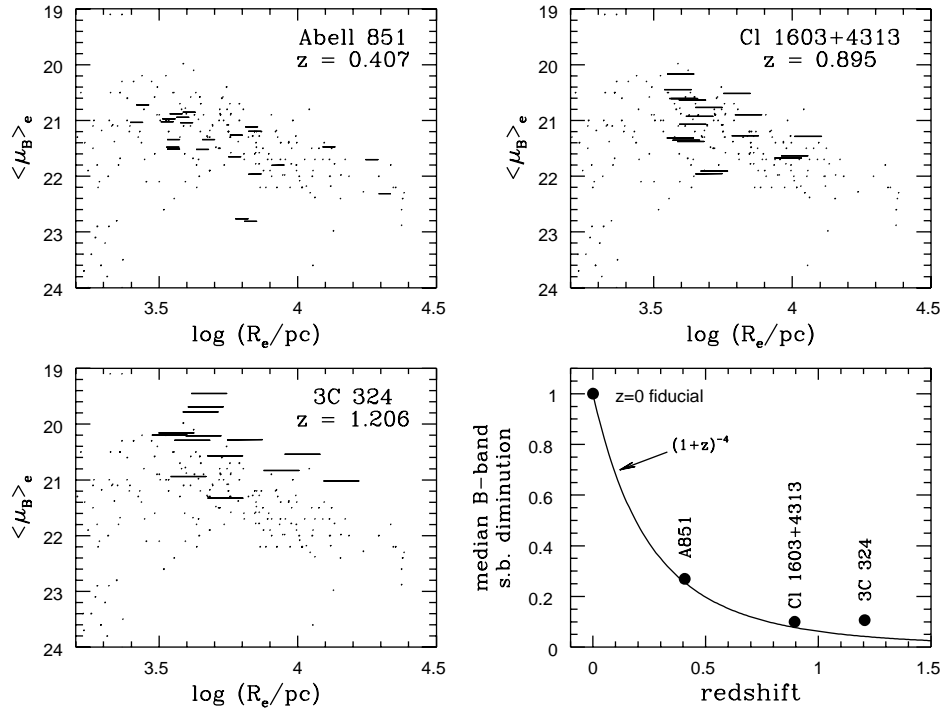

Figure 5. Sizes and rest–frame $B$ surface brightnesses of elliptical galaxies in high redshift clusters. The small dots are a local sample, while the dashes show the high–$z$ galaxies, with each dash connecting $R_e$ values for $q_0 = 0.5$ and 0. Cosmological dimming is illustrated in the lower right panel, which intercompares raw surface brightnesses from the three data sets (normalized to unity at $z = 0$).

## 4. Structural parameters, luminosity evolution, and the Tolman Test

The new *HST* data are good enough that we may start to make quantitative structural comparisons between high–$z$ ellipticals and their present–day counterparts. Ideally, one would like to measure the fundamental plane at high redshift. This must await high signal–to–noise spectroscopy for measuring velocity dispersions (cf. Franx 1993). In the meanwhile, we can consider projections of the fundamental plane which depend only on morphological/photometric parameters. One example is the observed anti–correlation between galaxy size and surface brightness (Kormendy 1977). Figure 5 plots effective radii $r_e$ and surface brightnesses $\langle\mu\rangle_e$ in the B–band for a sample of nearby elliptical galaxies compiled by Sandage & Perelmuter (1990). Data for high–$z$ cluster ellipticals measured from *HST* images are superimposed.

Two corrections are needed to transform the measured surface brightnesses to rest–frame $B$. One is the k–correction. For A851 and Cl 1603+43, this is small because the *HST* data were taken through the F702W and F814W filters respectively, bands whose effective wavelengths match that of the redshifted $B$–band reasonably well. For 3C 324 this term is larger, requiring correction from

rest–frame 3200Å to the $B$–band. This was made using a single–burst Bruzual & Charlot model spectrum with an age of 3.3 Gyr (see §3).

The second correction is the cosmological $(1+z)^4$ diminution (e.g. Tolman 1930), which becomes very large at high redshifts, making it possible to measure it directly. This "Tolman Test" has never been satisfactorily carried out for lack of an adequate surface brightness standard candle. Sandage & Perelmuter suggested using the surface brightness distribution of cluster ellipticals, and with *HST* data we can do this properly for the first time.

At $z = 0.4$ and 0.9, the cluster galaxies fall neatly onto the locus of present–day ellipticals. By $z = 1.2$ some deviation is seen toward higher surface brightnesses (or larger sizes). The lower right panel of figure 5 compares median surface brightness offsets from the normal Kormendy relation to the expected cosmological dimming. Reassuringly, the Tolman signal is strongly evident. The distant galaxies appear to lift above the expected curve, however, with the biggest departure ($\sim$ 1 mag) found at the highest redshift. This is presumably the signature of luminosity evolution. (The alternative, that the galaxies are evolving toward smaller sizes at a fixed surface brightness, seems dynamically implausible.) Indeed, about 1 mag of brightening is expected given the observed colors of the galaxies around 3C 324 (see §3). The greatest uncertainty comes from the k–correction for the 3C 324 galaxies. The adopted value is perhaps an upper limit – a model with a more protracted star–formation history would produce a flatter ultraviolet spectrum and reduce the inferred evolution somewhat.

## 5. Conclusion

The data presented here do not *prove* that all cluster ellipticals formed at high redshift and have evolved passively thereafter, but are consistent with such a scenario. The red, slowly evolving colors are well fit by single–burst models and suggest high formation redshifts. The small color scatter implies closely synchronized star formation histories, and the color–magnitude relation is preserved at least to $z = 0.9$, suggesting little differential evolution between bright and faint ellipticals. Remarkably little luminosity evolution is evident out to $z \approx 0.9$, perhaps even less than predicted by population synthesis models using a normal IMF. If galaxies in clusters transform themselves from blue disks to red E/S0s by mergers or other processes, they must do so without disturbing these tight relations. This is not to say that merging and cannibalism do not occur – figure 6 shows a striking example of "shells" in a dominant cluster elliptical at $z = 0.4$. But extensive star formation at late times seems incompatible with the photometric data on most cluster E/S0s.

It may be that studying only very rich clusters introduces some bias, reducing the apparent evolution of the elliptical *population* relative to that which holds for *individual* galaxies over a Hubble Time. In hierarchical theories of structure formation such as CDM, rich clusters originate from rare, high waves in the spectrum of mass fluctuations. Cluster ellipticals form early as biased peaks on top of those waves. By observing only the richest clusters, we may be selecting increasingly biased, rare environments at higher redshifts. The *local* epoch of spheroid collapse may have been earlier for higher redshift clusters. This would then have the effect of flattening the apparent degree of evolution.

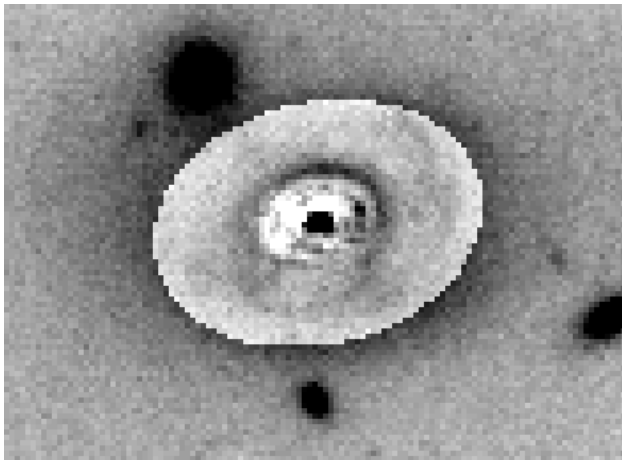

Figure 6.   *HST* image of the brightest elliptical (DG #311) in A851 at $z = 0.4$. A smooth $r^{1/4}$–law model has been subtracted to enhance the visibility of a bright shell–like structure in the galaxy envelope.

Eventually, data on high redshift ellipticals in other environments, i.e. groups and the field, can test this hypothesis. Data will pour in over the next few years as additional clusters are observed with *HST*, allowing us to intercompare galaxy properties at any particular redshift. The corresponding histories of ellipticals in low density environments should emerge from deep redshift surveys and *HST* imaging programs such as the Medium Deep Survey. This paper has presented data for a few points along the cosmic timeline – soon we should be able to fill in the redshift gaps and broaden the environmental spectrum. The existence of clusters such as that around 3C 324 gives hope that we may push the redshift limits higher, and thus closer to the earliest formation era.

**Acknowledgments.**   I thank my collaborators profusely, particularly Adam Stanford and Peter Eisenhardt, for their contributions and for endless discussions. §2 above relies heavily upon their work. I am also grateful to other *HST* investigators, particularly Alan Dressler and his collaborators, for obtaining some of the excellent data upon which I have drawn for this review. Finally, my thanks to the organizers for their hospitality, and for bringing me here to learn something about elliptical galaxies.

**Discussion**

*Nelson Caldwell*: You have lumped S0s with Es and conclude that both show such small color spread at high $z$ that they formed at the same, early epoch. Given that blue cluster galaxies at $z \sim 0.5$ have been shown to be spirals, and that these probably evolve into S0s, would you entertain the possibility that Es and bulges formed early, but disks, even in clusters, formed over a long time?

*Dickinson*: Perhaps, although it's not necessary that disks *formed* late – just that they somehow *disappeared* late in clusters. Naturally, the spirals had to go somewhere. S0s are one candidate, but recent simulations by Moore *et al.* suggest other possible fates. It would be interesting to look carefully at the redshift evolution of the S0–to–elliptical number ratio to test the spiral-→S0 picture.

*Dave Silva*: Lavery & Henry have claimed evidence for a higher rate of galaxy encounters and mergers in distant clusters. From your 3C 324 images, can you confirm this and say whether the trend extends to still higher redshifts?

*Dickinson*: Not yet, at least not without much more spectroscopy. Projection effects are horrendous at this redshift. We've recently gotten Keck spectra which show that one of the most spectacular "interacting groups" in our *HST* image is really a projection of galaxies with at least three different redshifts.

*Tim de Zeeuw*: A naive question: how much further into the IR can you go, in order to push this work to even higher redshifts?

*Dickinson*: Don't call that naive – it's my next observing proposal! I don't know the answer until we try. At some higher $z$ we may lose any unique red color signature if the gEs start getting much bluer. For $z \gtrsim 2.5$ even the $K$–band is getting awfully blue in the rest–frame, and observing faint galaxies at longer wavelengths may require IR satellites like SIRTF or ISO. On the other hand, Stanford, Elston and Eisenhardt have recently had some success with $L$–band imaging of clusters at $z \approx 0.4$, so who knows what will be possible someday?